\documentclass[conference]{IEEEtran}

\ifCLASSINFOpdf
\else
\fi

\usepackage[pdftex]{graphicx}
\usepackage{array}
\usepackage{multirow}
\usepackage{subfig}
\usepackage{float}
\usepackage{subfig}

\begin{document}
\setlength{\parskip}{1pt}
\title{A First Look at Mobile Intelligence: Architecture, Experimentation and Challenges}

\author{\IEEEauthorblockN{Ziyi Wang}
\IEEEauthorblockA{Tsinghua University}
\and
\IEEEauthorblockN{Yong Cui}
\IEEEauthorblockA{Tsinghua University}
\and
\IEEEauthorblockN{Zeqi Lai}
\IEEEauthorblockA{Tsinghua University}}

\maketitle

\begin{abstract}
Artificial intelligence (AI) technology makes mobile devices become intelligent objects which can learn and act automatically.
Although AI will bring great opportunities for mobile applications, little work has focused on the architecture and
the interaction with the cloud.
In this article, we present three existing architectures of mobile intelligence in detail and introduce its broad application prospects.
Furthermore, we conduct a series of experiments to evaluate the performance of the prevalent commercial applications and intelligent frameworks.
Our results show that there is a big gap between Quality of Experience (QoE) requirements and the status quo.
So far, we have seen only the tip of the iceberg.
We pose issues and challenges to advance the area of mobile intelligence and hope to pave the way for the forthcoming.
\footnote{\emph{This paper was accepted by the IEEE Network in May 2018 and will be published soon.}}
\end{abstract}

\section{introduction}
AI has recently attracted significant attention from both industry and academia,
as it gives the machine the ability to perceive its environment and take actions.
Specifically, it can extract high-level features from image, audio, or other signals automatically,
leading to a wide range of applications including computer vision, speech and natural language processing.
In the meantime, mobile devices have become both ubiquitous and increasingly powerful.
A large volume of multimedia data is being produced and released into mobile cellular networks \cite{liu2016device}.
Therefore, there is an increasing interest in applying AI to mobile environments.
Among existing mobile intelligent applications, Machine Learning (ML) is the most commonly-used technology.
Thus in this article, we focus on the intelligent applications based on ML.\\

Previous works on mobile intelligence have only focused on the hardware platforms or the software models.
Specifically, some teams are optimizing mobile hardware chips to support the operation of the ML model,
others try their best to build lightweight models without loss of learning performance.
However, there is scant research on the architecture choice of the mobile intelligent applications.
It is important to understand the existing architectures and optimize it from a more global perspective.
To fill this gap, in this article we present the first study on the architecture, experimentation and challenges
of mobile intelligence.\\

Firstly, we divide the existing intelligent applications into three
different architectures, namely cloud-based, local-based and partial offloading.
We provide a technical overview including the introduction of the system architecture, major components
and detailed functionalities. This architecture is applicable to all the mainstream ML models.
Some researchers have developed intelligent applications using local-based \cite{lane2015deepear}\cite{mcgraw2016personalized}
and others have adopted cloud-based \cite{peng2015deepcamera}\cite{zhang2017zipnet}.
There is also research work concerning combined models, such as making dynamic decisions on local-based or cloud-based \cite{ran2017delivering}.
Moreover, some researchers are exploring new architecture: partial offloading \cite{kang2017neurosurgeon}\cite{eshratifar2018jointdnn}.
On this basis, we propose three important QoE metrics to evaluate the performance of these mobile intelligent applications.\\

Around these metrics, we conduct some measurements on prevalent commercial applications and intelligent frameworks.
In the process of measuring Google Translate, we have selected two functions, namely Word Lens and Speech-to-speech translation,
which represent the local-based and cloud-based architectures respectively.
In the process of measuring TensorFlow's application programming interfaces (APIs), we have
developed two applications, namely TF-local-based and TF-cloud-based,
which represent the local-based and cloud-based architectures respectively.
Using both black-box testing and white-box testing, we get important metrics such as latency, CPU/RAM utilization and discharge rate.
For the data obtained, we sort them out and find the mean and standard deviation.
We conclude all experiment results and give some analysis.
We find that there is indeed a big gap between QoE requirements and the status quo.
Furthermore, we conduct a measurement study on partial offloading architecture using Inception-v3 model \cite{szegedy2016rethinking}.
We find that the best partition point for latency is closely related to network bandwidth rate and the computational capability of the mobile device.\\

Since there are many difficulties and challenges on the way to mobile intelligence,
we propose the key challenges which are most likely to appear and give some insights for future improvement.
Specifically, we consider unstable network conditions, considerable energy consuming, privacy disclosure,
increasing model complexity and coarse-grained partition of the inference process.
To the best of our knowledge, this is the first article that provides a wide overview and experimental evaluation
for the existing architectures of the mobile intelligent applications.\\

\section{Architecture}

\begin{figure*}
\centering
\includegraphics[width=18cm,height=8.7cm]{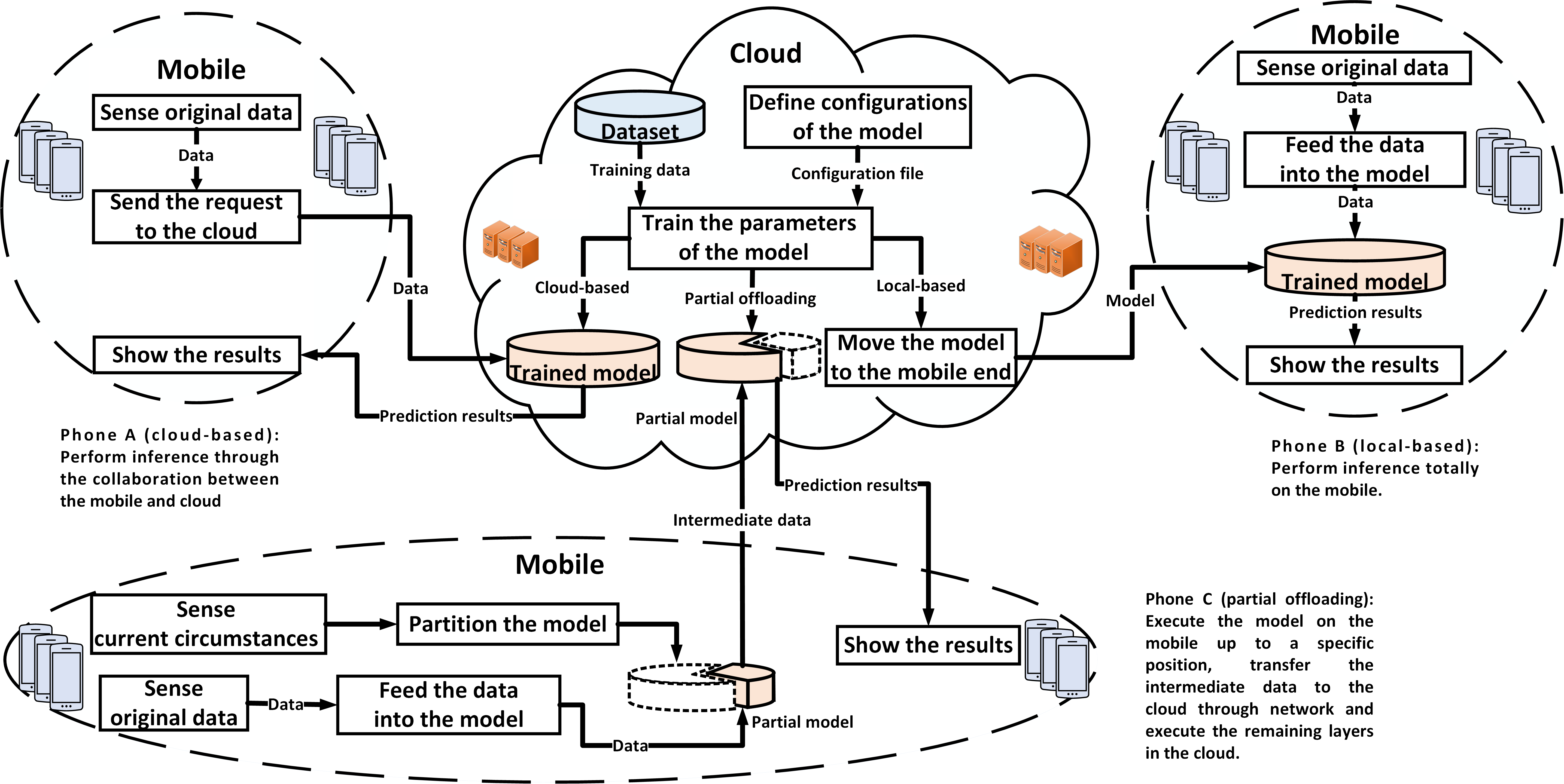}
\caption{Mobile intelligence architecture: cloud-based, local-based and partial offloading.}
\label{Figure 1}
\end{figure*}

ML models are particularly well suited for performing perceptual tasks,
which can sense, learn from and respond to their environment.
Depending on the location of these trained models, we divide the existing applications into three different architectures, namely cloud-based, local-based and partial offloading,
which are illustrated in Fig. \ref{Figure 1}. Two major components can be identified in this figure: the mobile client and the cloud server.
We first introduce the detailed functionalities of these two components.\\


\textbf{Mobile client:} Mobile client receives input signals and preprocesses them locally.
Then the mobile sends them either to the cloud's ML model, or to the local model.
After processing, the mobile client obtains the prediction results and presents the information to the user.\\

\textbf{Cloud server:} The cloud server has abundant computing resources such as CPU, GPU and TPU, by which the cloud server can
complete the training of ML model.
In order to train it, we need to provide the cloud server with the training data and configuration files of the related models.
The cloud can also continue to carry well-trained models and provide web APIs to help inference processing.\\

As shown in Fig. \ref{Figure 1}, Phone A, B and C represent three typical architectures respectively. Here we briefly describe their workflow
and their advantages and disadvantages.\\

Phone A is the cloud-based one, which means the mobile client and cloud server work together to make prediction including a training process and an inference process.
When training is done on the server, the cloud server obtains the learned parameters for the model.
Then we can put the trained model on the server and publish web APIs which mobile devices can use.
Since the model is on the server, it is easy to port the application to different platforms.
However, inference depends on network and cannot be done locally on the device.\\

Phone B is the local-based one, which means only the mobile makes prediction.
We put the trained model into mobile devices and inference locally.
We don't need to ask the server over the network during the inference process.
It can be faster and more reliable.
However, it requires large amounts of CPU and RAM resources on the mobile.\\

Phone C represents the partial offloading architecture, which is a more flexible and dynamic one.
The model is composed of many abstract layers.
On one hand, the mobile client partitions the model according to the current circumstances, including network condition, mobile capability and server load.
On the other hand, it executes the model up to a specific layer and transfers the intermediate data to the cloud through network.
Then the cloud server executes the remaining layers and sends the prediction results back to the mobile client.
This architecture would be more appealing when mobile applications are becoming more and more intelligent.\\

The architecture above is universal to which the mainstream ML models are all applicable, such as Deep Neural Network (DNN),
Reinforcement Learning (RL) models and Generative Adversarial Network (GAN).
The only thing we need to do is to make the corresponding replacement for the specific model.\\

Based on these three architectures, we have seen diverse mobile intelligent assistants such as Google Home, Apple Siri and Microsoft Cortana.
All of them use accurate and complex ML technologies to process voice signals.
In order to better depict the user experience of these mobile intelligent applications, we introduce three QoE metrics.\\

\textbf{Latency:} Latency refers to the time that elapses between the user's request and the prediction results,
including pre-processing, model operation and post-processing.
For some real-time interactive intelligent applications, such as mobile Virtual Reality (VR),
they require 14ms latency and 60FPS (the phone display refresh rate) \cite{lai2017furion}.
For cloud gaming providers, interaction latency must be kept as short as possible in order to provide a rich experience to cloud gaming players \cite{shea2013cloud}.\\

\textbf{Accuracy:} Accuracy refers to the ratio of the number of samples that get the correct results to the total number of samples,
which can be used to measure the performance of the model.
For some applications requiring a high level of security, such as autonomous driving and road navigation, they require ultrahigh accuracy.
Inaccuracy of any prediction result will be life-threatening.
Some researchers have proposed that a well-trained DNN can predict the steering angle with an accuracy close to that of a human driver \cite{bojarski2016end}.\\

\textbf{Energy:} Mobile devices are energy-constrained. However, running these complex models
can introduce considerable computing and communication overhead.
Although mobile intelligent applications are very attractive to users,
they will most likely choose not to use them if the energy consumption is huge.
Therefore, energy efficiency is a desired goal in these mobile intelligent applications.\\

\section{experimentation}

There have been a lot of daily-used commercial mobile intelligent applications, such as Google Translate \cite{wu2016google}.
In addition, many effective open-source libraries and frameworks have also appeared, such as Tensorflow,
which provides convenience for developing intelligent applications on mobile devices.
We conduct a measurement study to quantitatively describe their QoE level.
Specifically, we measure from two perspectives: commercial mobile intelligent applications and mobile intelligent frameworks.
Furthermore, We also measure the QoE on the partial offloading architecture based on Inception-v3 model \cite{szegedy2016rethinking}.
We run the applications on a Nexus 6P smartphone. The data is sent to the cloud over the wireless network.\\

\subsection{Measurement on commercial mobile intelligent applications}
\begin{figure}[H]
\raggedleft
\includegraphics[width=90mm]{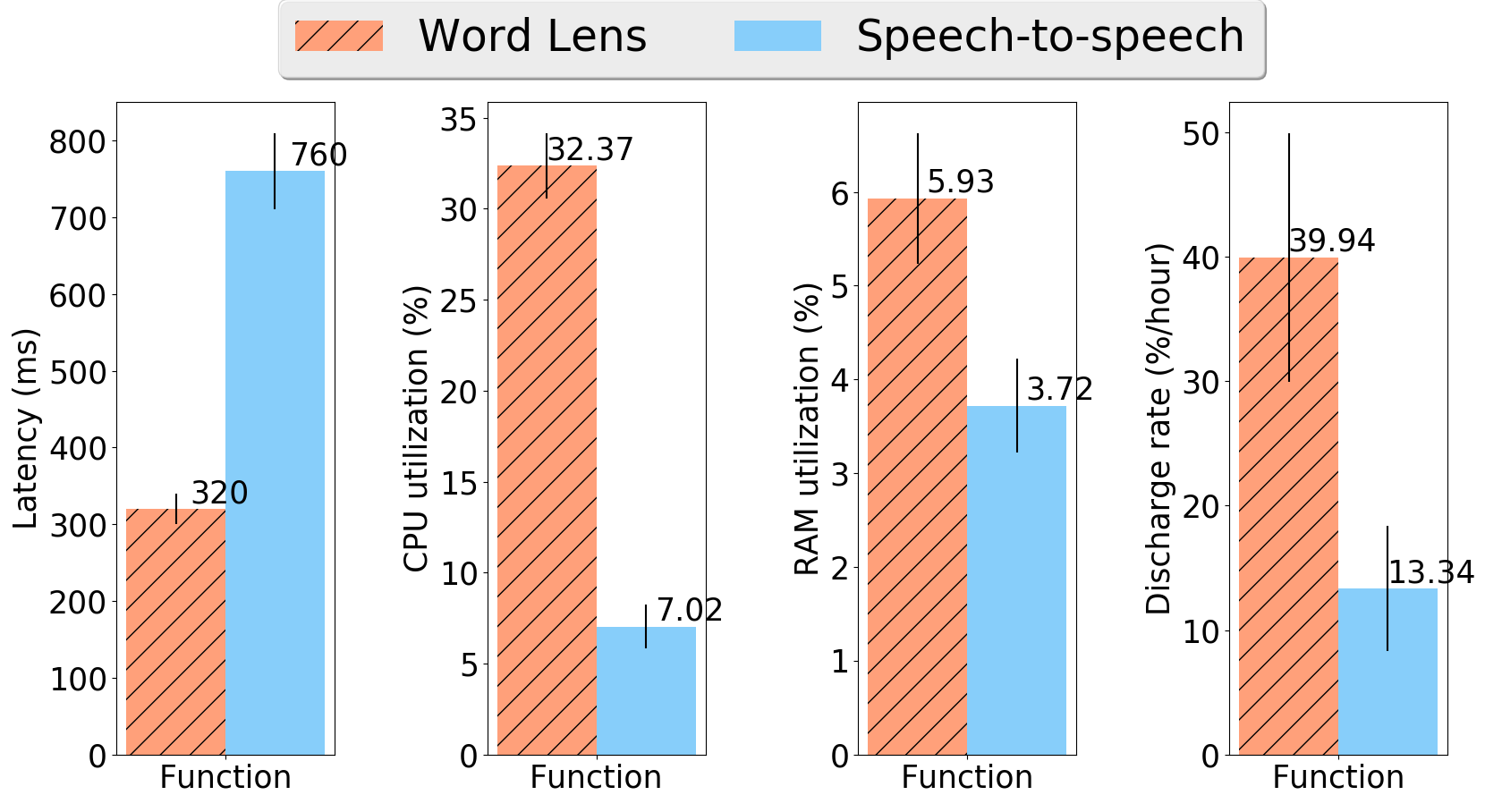}
\caption{Latency, CPU/RAM utilization and discharge rate of Word Lens and speech-to-speech translation.}
\label{Figure 2}
\end{figure}

We first measure Google Translate, one of the most commonly used mobile application.
When using its speech-to-speech translation function, we need to connect the Internet.
Hence, it belongs to the cloud-based architecture.
However, Google Translate's augmented reality feature, Word Lens, is done through offline language packs.
Consequently, it belongs to the local-based architecture.
Since the source code for the app is not public, we conduct a black box test by recording video.
Specifically, we collect 100 images and 100 sentences in English,
which are transmitted to the mobile application (Google Translate) in the form of image and voice respectively.
In the process of translating these sentences from English into Chinese, we record it into videos.
Then we analyze the video frame by frame and calculate the latency of processing each image or voice.
As for the CPU and RAM utilization, we use Emmagee software, which is a simple and easy-to-use Android performance monitoring tool.
Users can configure monitoring frequency and get performance statistics eventually.
What's more, we leverage the Google Battery Historian tool to inspect the discharge rate of the Android device over time.
For the data obtained, we sort them out and find the mean and standard deviation, as shown in the Fig. \ref{Figure 2}.\\

From the measurement results, we observe that the Word Lens function achieves lower latency, higher utilization rate of CPU/RAM and higher discharge rate.
Since it computes locally based on offline language packs, it is faster but more resource-consuming.
On the contrary, the speech-to-speech function has larger latency, lower utilization rate of CPU/RAM and lower discharge rate.
Since it sends voice to the cloud for processing, the network round-trip latency is larger while the local CPU/RAM resource utilization and discharge rate of mobile device is lower.
After more in-depth analysis, we find that the latency of the two functions are in the hundred-millisecond level, which is relatively large.
In the measurement of Word Lens, we find if we move the smartphone in real time, it can not process immediately
to give the right results and it seems to be stalling.
In addition, this function only provides accurate translation for short and simple sentences.
Once complex texts appear, the accuracy rate is greatly reduced.
Worse still, some words are translated while others are not, which seriously affects the user experience.
What's more, CPU utilization of this function has reached 32.37\% and discharge rate has reached 39.94\% per hour, leading to high workload and energy consumption of smartphone.
In the measurement of speech-to-speech, we find that although the CPU/RAM resource utilization and discharge rate is lower, the latency is larger.
When we gradually weaken the wireless network, the latency can reach even few seconds, which is unbearable.
\\

\subsection{Measurement on mobile intelligent frameworks}
\begin{figure}[H]
\centering
\includegraphics[width=90mm]{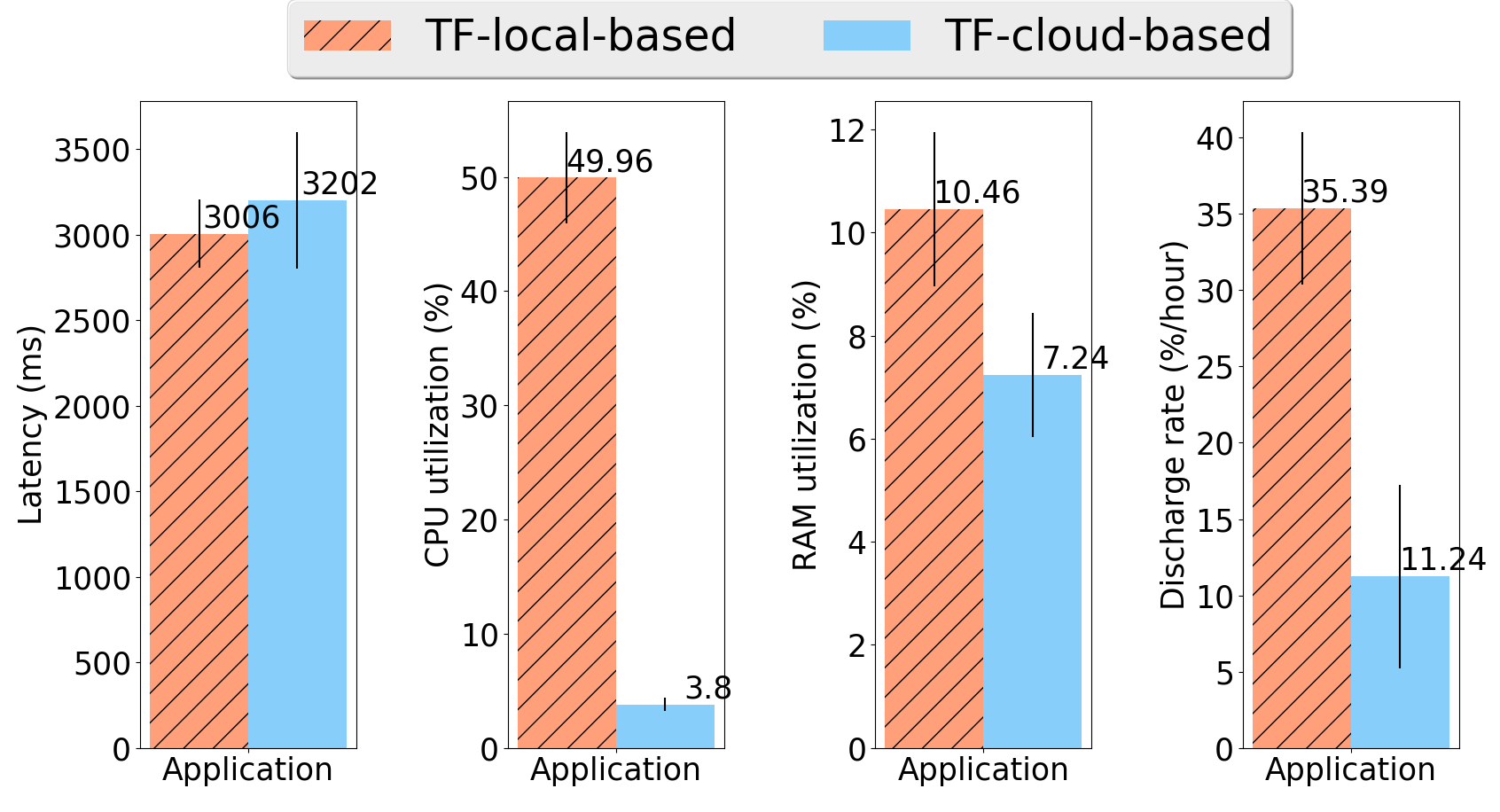}
\caption{Latency, CPU/RAM utilization and discharge rate of TF-local-based and TF-cloud-based.}
\label{Figure 3}
\end{figure}

TensorFlow is one of the most prevalent frameworks in the deep learning ecosystem.
It provides an inference interface which can be called to complete the entire neural network processing including input, running and output.
In order to measure its performance, we develop two applications which can classify camera image
based on the two kinds of architectures.
We call them TF-local-based and TF-cloud-based respectively.
TF-local-based can classify image and display the top results in an overlay on the camera image.
It runs the neural network totally on the mobile device.
In contrast, TF-cloud-based is a client-server architecture.
We first need to start a Flask web server preparing to receive the mobile's request.
When the mobile device captures an image, the application will send it to the server through the network.
The server receives the image and runs the neural network model to get the final results.
The top classification results will be sent back to the mobile edge through the network and
presented to the user.
We use Inception-v3 model trained on the ImageNet Large Visual Recognition Challenge dataset for both applications.
The model can differentiate between 1,000 different classes.
During the measurement, we collect 100 images from test set and transmit them to these two applications respectively.
Since we have source code for both applications, we measure the latency by inserting timestamps into the code.
Latency refers to the time that elapses between the image request and the prediction result.
For the CPU and RAM utilization measurement, we still use Emmagee software.
For the battery energy measurement, we still use Google Battery Historian.
We also compare latency, CPU/RAM utilization and discharge rate between them, which are illustrated in Fig. \ref{Figure 3}.\\

From the experimental results, we can find that the latencies of both applications are more than 3000ms,
under which condition real-time object classification is not applicable.
More seriously, TF-local-based's CPU and RAM utilization reach 49.96\% and 10.46\%, respectively, which seriously affects the normal operation of the smartphone.
What's more, its discharge rate is about 35.39\% per hour, which means this application can only last for 2.83 hours.\\

Combining all the measurement results, we can find that
existing cloud-based and local-based solutions do not meet the needs of users.
Although ML brings intelligence to mobile applications, there still exist hundreds of milliseconds or even seconds in terms of latency.
CPU and RAM utilization is excessively high and the corresponding energy consumption is increasing.
In addition, accuracy of the processing results is far from satisfactory.
Hence, there is indeed a big gap between QoE requirements and the status quo.\\

\subsection{Measurement on partial offloading architecture}
\begin{figure*}[!htb]
\centering
\subfloat[Pixel, Bandwidth=0.2MB/s]{\includegraphics[width=0.5\textwidth]{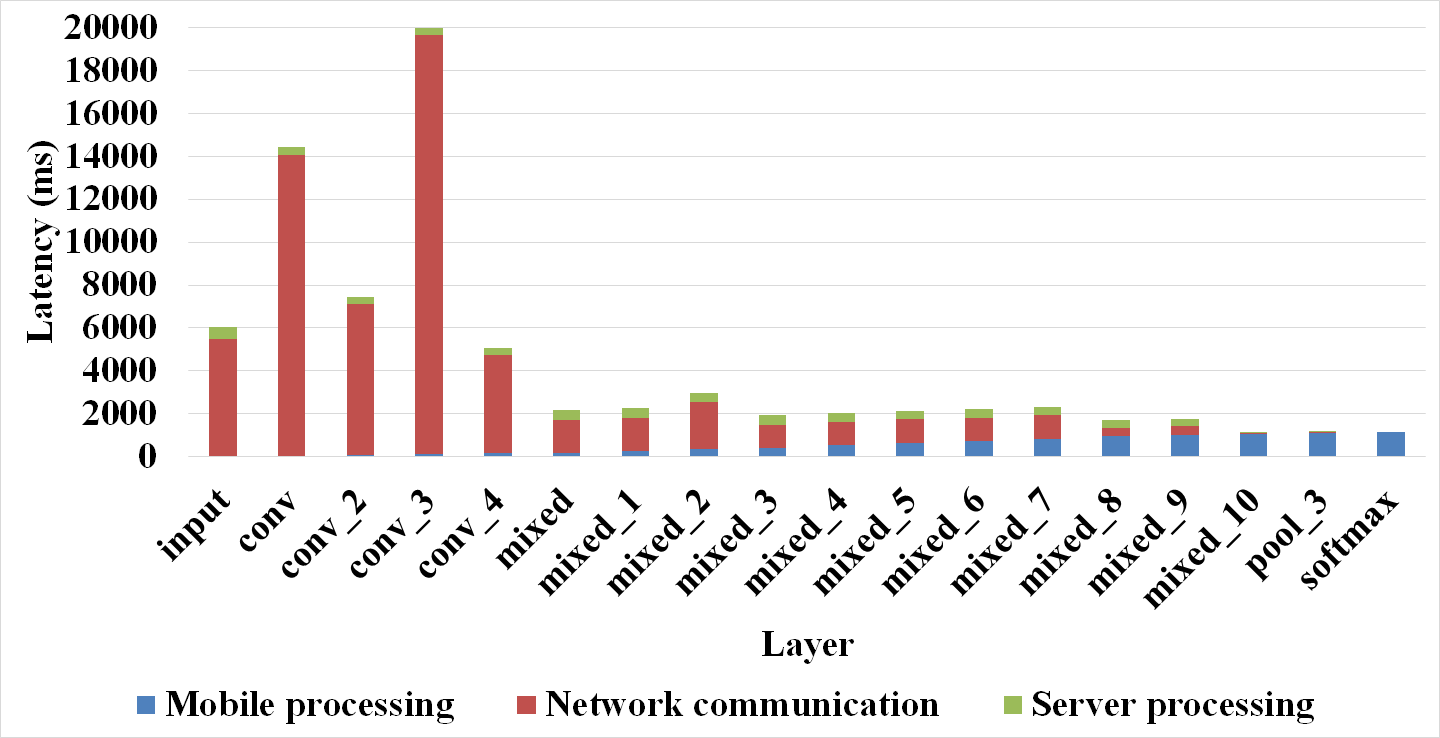}}
\subfloat[Nexus 6P, Bandwidth=0.2MB/s]{\includegraphics[width=0.5\textwidth]{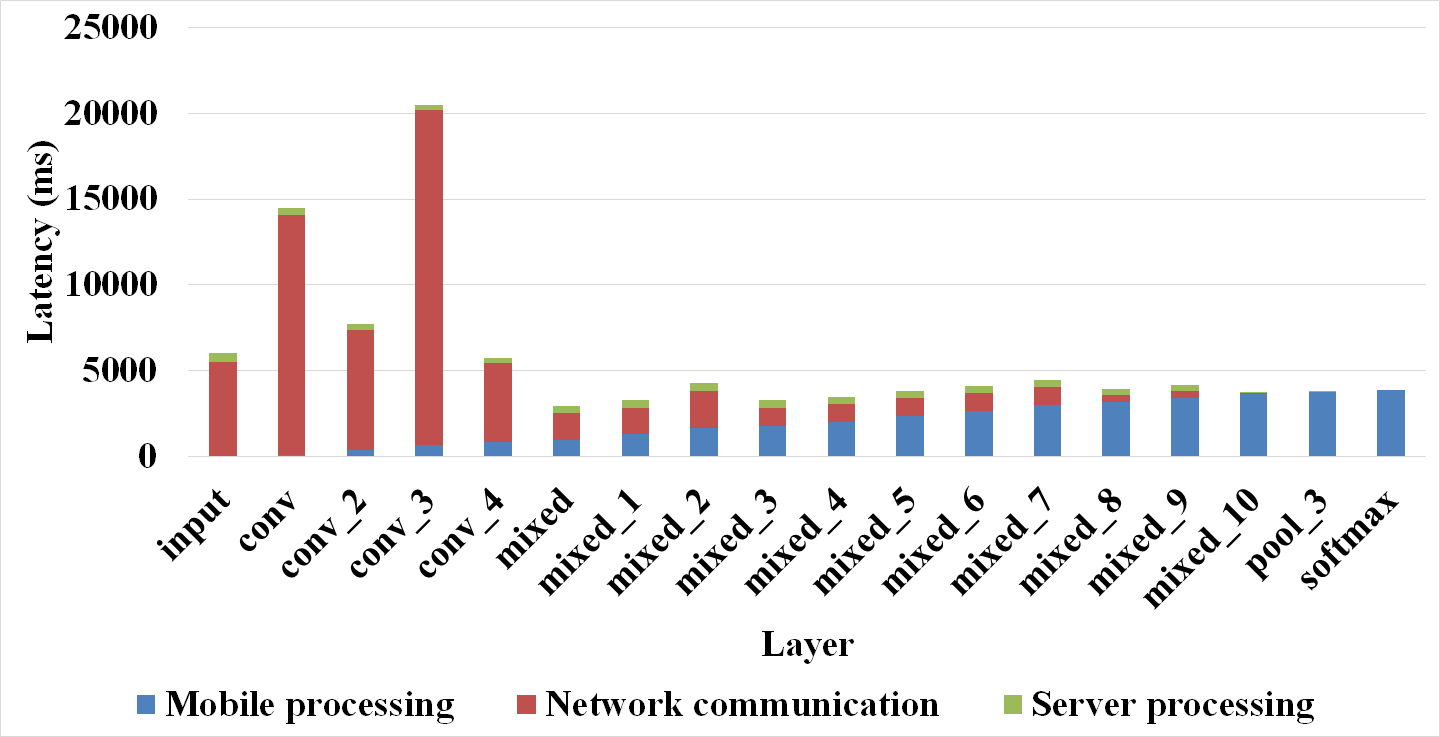}}\\
\subfloat[Pixel, Bandwidth=1MB/s]{\includegraphics[width=0.5\textwidth]{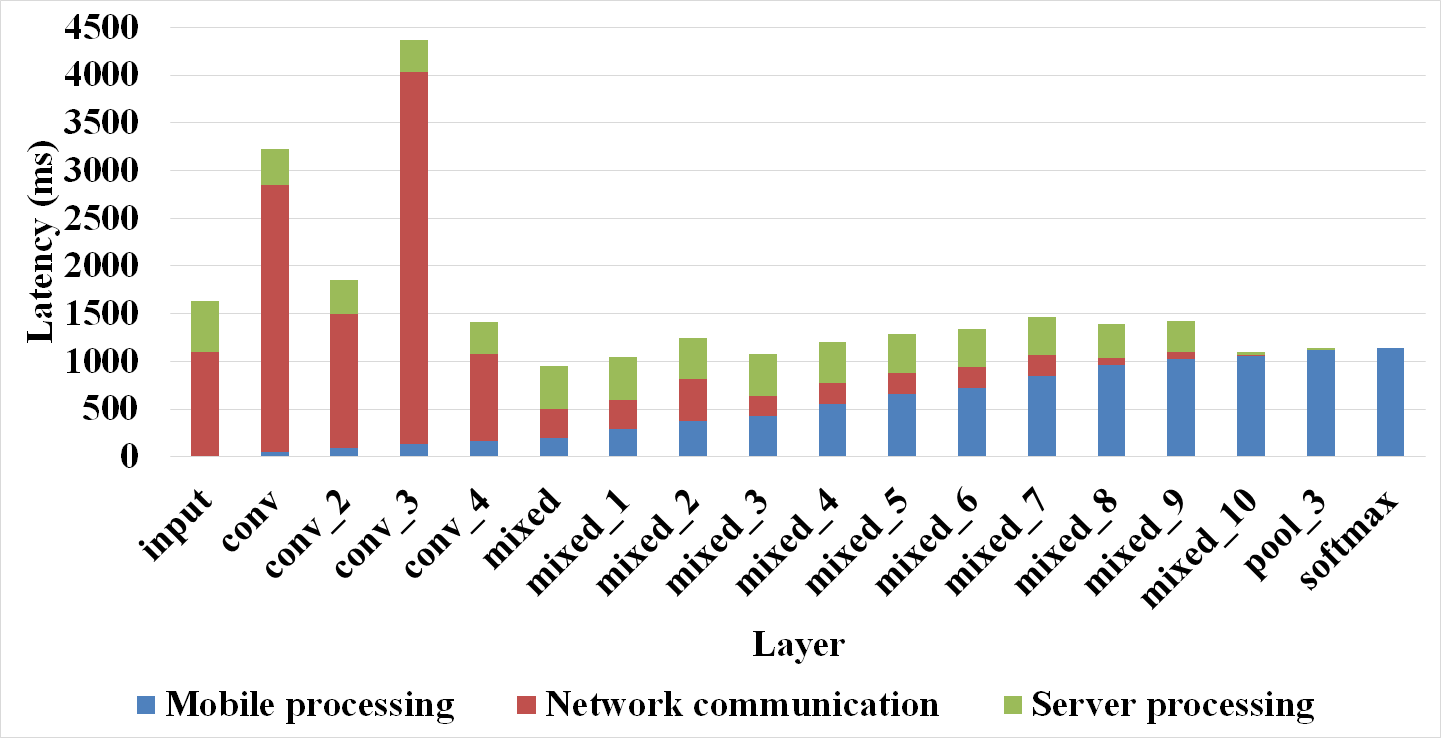}}
\subfloat[Nexus 6P, Bandwidth=1MB/s]{\includegraphics[width=0.5\textwidth]{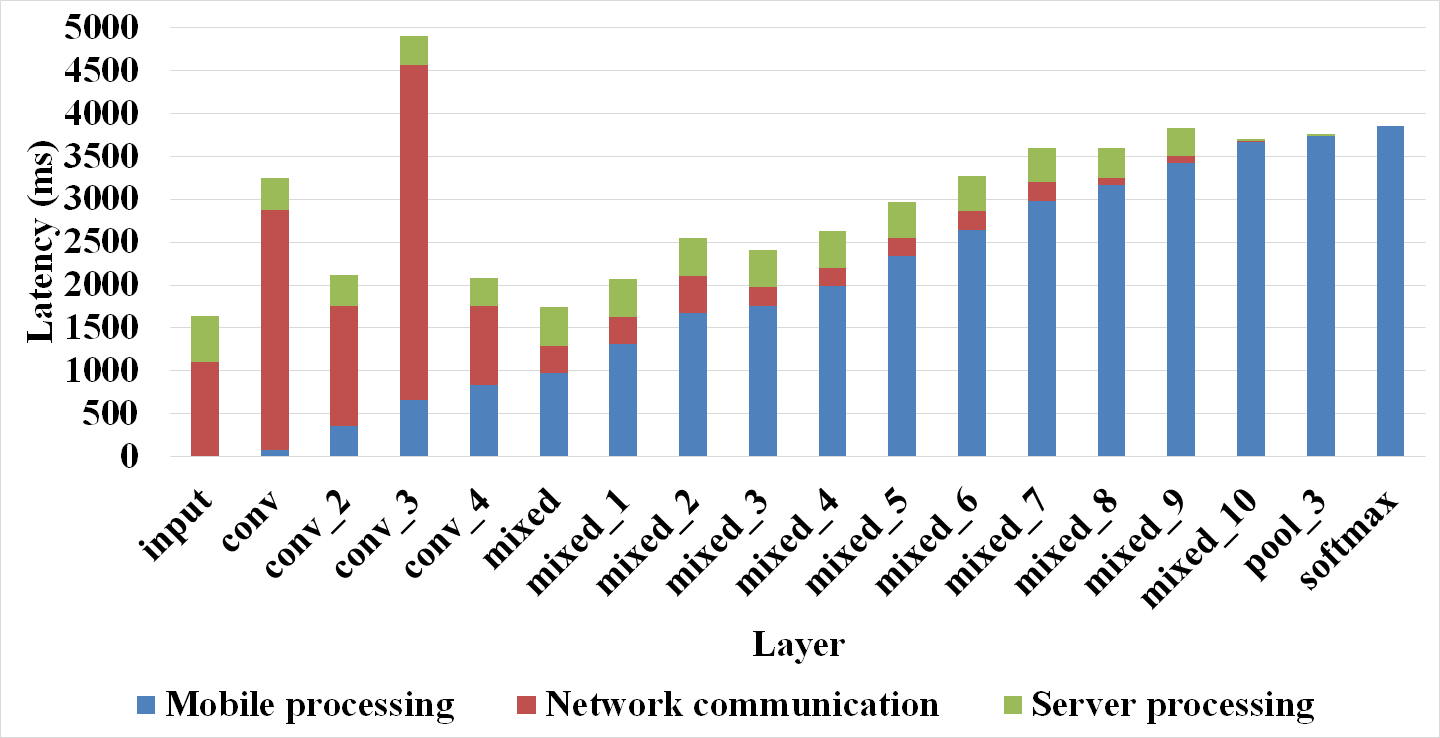}}\\
\subfloat[Pixel, Bandwidth=5MB/s]{\includegraphics[width=0.5\textwidth]{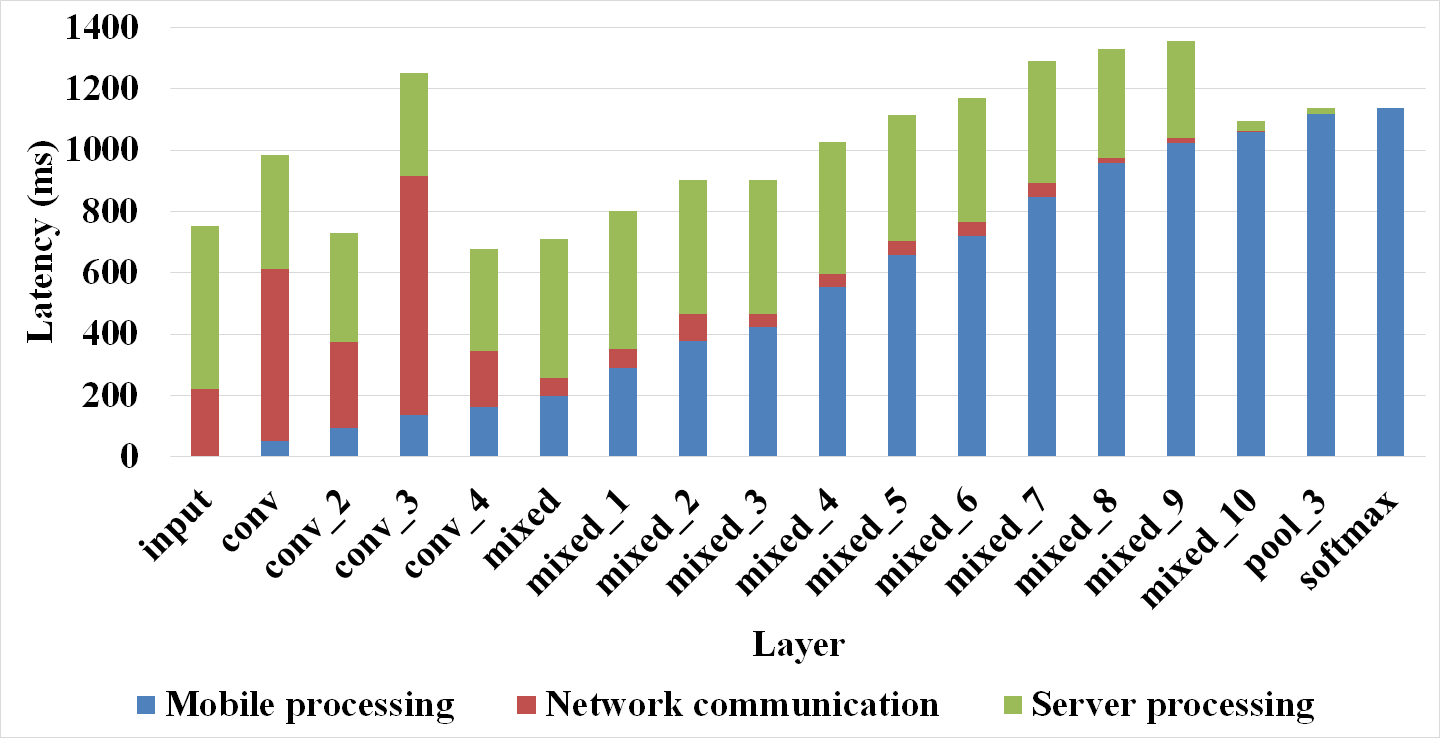}}
\subfloat[Nexus 6P, Bandwidth=5MB/s]{\includegraphics[width=0.5\textwidth]{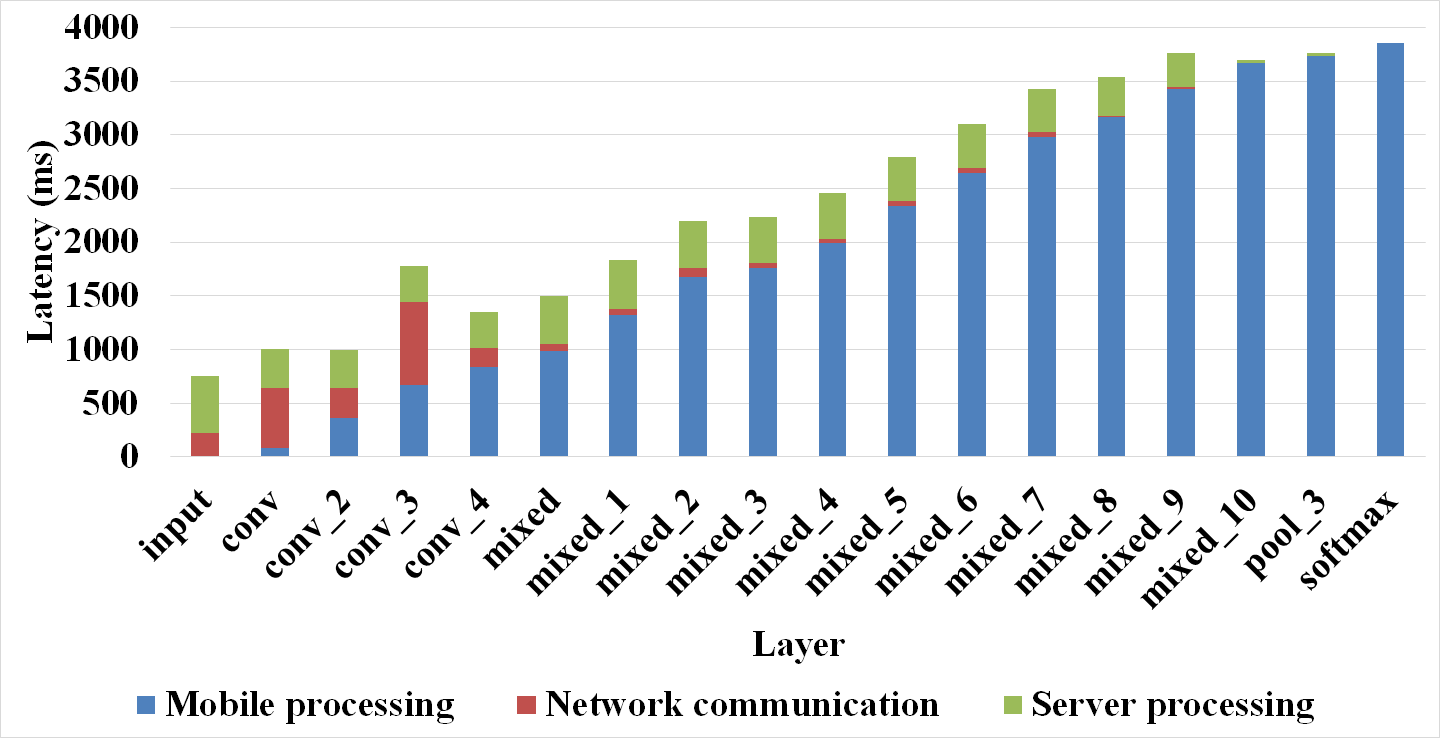}}\\
\caption{End-to-end latency when choosing different partition points with different mobile devices and network conditions.}
\label{fig:geo_distribution}
\vspace{-0.2in}
\end{figure*}

Since both cloud-based and local-based architectures fail to meet the requirements, we make some measurements on a new architecture: partial offloading.
We develop an application based on Tensorflow which can classify the images captured by the phone camera.
We partition Inception-v3 model at the layer granularity.
Specifically, we set each layer as a partition point.
For the given partition point, mobile-end executes the computation up to it and transfers intermediate data to the cloud.
Next, cloud executes the remaining layers and transfers the prediction results back to the mobile-end.
For each partition method, we send 100 test images to the application and compute the average latency.
We make experiments under different network bandwidth (0.2, 1 and 5MB/s) and different mobile phones (Pixel and Nexus 6P) which represent various computation capabilities.
Since we have source code for both applications, we break down the end-to-end processing latency, including mobile processing, network communication and server processing.
The results are shown in Fig. \ref{fig:geo_distribution}.
Each bar represents the end-to-end latency for a specific partition way.
The leftmost bar represents the cloud-based architecture while the rightmost bar belongs to the local-based architecture.\\

From the results, we can find that every layer has a totally different computational capacity.
The best partition point for latency is different under different circumstances,
which is closely related to network bandwidth rate and the computational capability of the mobile device.
We can also find that these existing best results are still high and far from meeting the users' need for latency.\\

\section{challenges}

Since there is a huge gap between QoE requirements and the status quo, we should make every effort to bridge it.
However, during this process we may face many challenges as we highlight in this section.\\

\textbf{Network condition is unstable, unsatisfied and unpredictable.}
Network condition is constantly changing and it is difficult to select a fixed formula to characterize it.
In addition, for some mobile VR applications, the existing network situation is far away from the QoE requirements.
Therefore, it is a challenge to dynamically assign tasks between the mobile and the cloud according to diverse network conditions.
A relatively simple method is that we use some regression models to predict the current wireless network
conditions based on some real-time probe data.\\

\textbf{Either local computing or communicating with cloud will consume considerable energy.}
The successful operation of mobile assistant requires a lot of computation and communication overhead.
To solve this problem, we need to propose a more efficient mobile energy-saving mechanism.
A viable solution is to develop an energy model tool to record data flow and energy flow.
It tells us how much energy the model consumes and where the bottleneck exists.
Then we can use such information to design new energy-efficient models or to optimize the existing models.\\

\textbf{Cooperation with the cloud will inevitably produce the issue of privacy disclosure.}
The data collected by the mobile devices can be very sensitive and private.
Uploading these information onto the cloud without any preprocessing constitutes a great danger to an individual's privacy.
In the future, users may have the choice to use a different method to process these data (local-based, cloud-based or partial offloading),
depending on which option best suits the situation.\\

\textbf{Model complexity and data size are increasing.}
Take the example of deep learning. The models are becoming more and more complex, with the number of parameters and layers increasing significantly.
Although this change improves the performance of models, it also brings new challenges in adapting resource-constrained mobile to
these advanced models.
To deal with this challenge, some teams provide hardware solutions.
For example, Huawei's new flagship Kirin 970 is Huawei's first mobile AI computing platform featuring
a dedicated Neural Processing Unit (NPU). This chip can perform the same AI computing tasks faster
and with less power.
In the meantime, some teams are developing extending software frameworks for the mobile.
For example, Google has announced Tensorflow Lite which is a lightweight solution for mobile and embedded
devices. It can also support hardware acceleration with the Android Neural Networks API.\\

\textbf{Current partition of the inference process is still coarse-grained.}
Actually, many models can be split into different kinds of modules which are respectively responsible for different functions.
In addition, distribution of latency varies a lot and is closely related to the corresponding workload.
For example, DeepMon \cite{huynh2017deepmon} indicates that the convolutional layers dominate the execution cycles in the VGG-VeryDeep-16 and YOLO model.
DeepEye \cite{mathur2017deepeye} demonstrates that the loading of fully-connected layers is the most time-consuming task across 8 different models.
Neurosurgeon \cite{kang2017neurosurgeon} indicates that for AlexNet, VGG and DeepFace, convolution layers are most time-consuming; for MNIST,
fully-connected layers are most time-consuming; for Kaldi and SENNA, layers of the model incur similar latency.
Faced with this situation, we should propose a deep integration architecture between mobile and cloud, which splits the functional
modules intelligently according to different workloads, models, network conditions and server loads.

\section{conclusion}
Since there will be more and more applications implemented with ML technology on the mobile, understanding the
existing architectures of the mobile intelligent applications is significant for both industry and academia.
In this article, we present a thorough overview of the mobile intelligence by introducing its architectures, components and functionalities
followed by an experimental study that evaluates the prevalent commercial applications and intelligent frameworks.
All tested services suffer performance limitations.
Our results show that there is a big gap between QoE requirements and the status quo.
Finally, we conclude experiment results and propose some challenges.
To the best of our knowledge, this is the first article that provides a wide overview and experimental evaluation
for the existing architectures of the mobile intelligent applications.
As for future work, we intend to do more detailed measurements and identify the bottleneck and propose advanced mobile intelligence architectures.



\bibliographystyle{unsrt}
\bibliography{ref}
\end{document}